\documentclass[%
reprint,
a4paper,
preprintnumbers,
aps,
showpacs,
prd]%
{revtex4-1}

\usepackage{epsfig}
\usepackage{epstopdf}
\usepackage[sumlimits,intlimits,namelimits]{amsmath}
\usepackage[mathscr]{euscript}
\usepackage{dsfont}

\usepackage[english]{babel}

\usepackage{graphicx}
\usepackage{subfigure}  

\usepackage{hyperref}

 \def\lag{\mathscr{L}}

\def\mws{M_W^2}

\newcommand\Op[2]{\mathcal{O}^{#1}_{#2}}
\newcommand\will[2]{\mathcal{C}^{#1}_{#2}}
\newcommand\ten[3]{#1^{#2}_{#3}}
\newcommand\coupL[2]{L^{#1}_{#2}}
\newcommand\coupR[2]{R^{#1}_{#2}}

\def\thdm{{\rm{2HDM}}}
\def\brf{\mathscr B}

\def\xq{x_q}
\def\xw{x_W}

\makeatletter
\def\fmslash{\@ifnextchar[{\fmsl@sh}{\fmsl@sh[0mu]}}
\def\fmsl@sh[#1]#2{%
  \mathchoice
    {\@fmsl@sh\displaystyle{#1}{#2}}%
    {\@fmsl@sh\textstyle{#1}{#2}}%
    {\@fmsl@sh\scriptstyle{#1}{#2}}%
    {\@fmsl@sh\scriptscriptstyle{#1}{#2}}}
\def\@fmsl@sh#1#2#3{\m@th\ooalign{$\hfil#1\mkern#2/\hfil$\crcr$#1#3$}}
\makeatother

\begin{document}
\title{Minimal Flavour Violation and Anomalous Top Decays}

\author{Sven \surname{Faller}}
\email[e-mail: ]{faller@physik.uni-siegen.de}
\affiliation{Theoretische Physik 1, Naturwissenschaftlich-Technische Fakult\"at, Universit\"at Siegen, D-57068 Siegen, Germany}
\author{Stefan \surname{Gadatsch}}
\email[e-mail: ]{gadatsch@nikhef.nl}
\affiliation{Nikhef, National Institute for Subatomatic Physics\\ P.O. Box 41882, 1009 Amsterdam, Netherlands}
\affiliation{CERN, CH-1211 Geneva 23, Switzerland}
\author{Thomas \surname{Mannel}}
\email[e-mail: ]{mannel@physik.uni-siegen.de}
\affiliation{Theoretische Physik 1, Naturwissenschaftlich-Technische Fakult\"at, Universit\"at Siegen, D-57068 Siegen, Germany}

\pacs{12.60.Fr,14.65.Ha,11.30.Hv}
\date{\today}
\preprint{SI-HEP-2012-09}
\preprint{Nikhef-2012-008}

\begin{abstract}
Top quark physics at the LHC may open a window to physics
beyond the standard model and even lead us to an understanding
of the phenomenon ``flavour''.
However, current flavour data is a strong hint
that no ``new physics'' with a generic flavour structure
can be expected in the TeV scale.
In turn, if there is ``new physics'' at the TeV scale,
it must be ``minimally flavour violating''.
This has become a widely accepted assumption for 
``new physics'' models.
In this paper we propose a model-independent scheme 
to test minimal flavour violation
for the anomalous charged $Wtq$,
$q\in\{d,s,b\}$,
and flavour-changing $Vtq$,
$q\in\{u,c\}$ and $V\in\{Z,\gamma,g\}$ couplings
within an effective field theory framework,
i.e., in a model-independent way.
We perform a spurion analysis of our effective field theory approach
and calculate the decay rates for the anomalous top-quark decays
in terms of the effective couplings for different helicities
by using a two-Higgs doublet model of type II,
under the assumption that the top-quark is produced at a high-energy collision
and decays as a quasi-free particle.
\end{abstract}
\maketitle

\section{Introduction}
Due to its large mass,
$m_t = 173.2\pm0.9$~GeV \cite{Lancaster:2011wr},
the top-quark seems to play a special role,
and therefore top-quark physics
may give hints to physics beyond the standard model (SM).
Assuming that the Higgs mechanism is indeed the way particles obtain their masses,
the top-quark is the only quark with a Yukawa coupling of a ``natural'' size,
i.e. a coupling of order unity.
Thus it may in particular lead us to an understanding of the phenomenon ``flavour'',
which in the SM is encoded in the Yukawa couplings. 

Given the fact that up to now we do not have any hint
at a specific model for ``new physics'' (NP),
and that the LHC \cite{Evans:2008zzb}
will produce a large amount of top-quarks \cite{Beneke:2000hk,Bernreuther:2008ju},
it is be desirable to have an approach
which is as model independent as possible to analyse the data.
Hence we will not pick a specific model here;
rather we shall refer to an effective theory description of possible NP.
This approach is well known and we shall gather
the necessary relations in the next section. 

However, by including up to dimension-6 operators in this approach
a large number of unknown couplings appear,
which are \textit{a priori} unconstrained.
From present data we may obtain limits on certain couplings
which have to be included in the analysis.
In particular, flavour physics rules out
a generic flavour structure for the dimension-6 operators,
and the flavour constraints can be readily incorporated
by the assumption of minimal flavour violation (MFV).   

MFV has become a popular assumption
to avoid flavour constraints in many new physics models.
On the other hand, it is important to test,
whether a hint at NP is compatible with MFV or not.
Hence it is desirable to have the possibility to test the MFV hypothesis
without referring to a specific new physics model,
in which case one can only make use of the effective theory approach. 

There is already a large number of analyses
of anomalous top couplings in the literature,
some recent work can be found in
Refs.~\cite{Willenbrock:2012br,Zhang:2012cd,Zhang:2010dr,AguilarSaavedra:2004wm}. 
However, these papers either deal with flavour-diagonal anomalous couplings
or do not take into account the constraints from MFV.  
In the present paper we propose a way
to perform a test of the MFV hypothesis in
anomalous flavour-changing top couplings.
The basic idea is to make use of the MFV constraints
on the flavour structure of the $t \to qV$ couplings,
where $q$ is a quark and $V$ a gauge boson.
However, this is still not restrictive enough to allow for a simple analysis 
and hence we will be forced to make additional assumptions
which we shall keep as simple as possible. 

The paper is organised as follows.
In the next section we collect the relevant dimension-6 operators
for an effective description at the top-mass scale.
In Sec.~\ref{sec:MFV} we give the MFV relations
between the coupling constants of these operators
and project out the relevant operators for charged- and neutral-current top decays.
In Sec.~\ref{sec:DRTop} we compute the rates for
top decays into lighter quarks under the emission of gluons,
photons and weak bosons and derive relations between different decay rates
which may serve as a test of MFV and conclude in Sec.~\ref{sec:Conc}.   
\section{Effective Approach with Two Higgs Doublets} \label{sec:EA-2HD}
We consider the SM as the dimension-4 part of an effective theory;
hence physics at a large scale $\Lambda$ beyond the SM manifests itself
through the presence of higher-dimensional operators,
suppressed by powers of the scale $\Lambda$. 
This is generically true for any new physics model
with degrees of freedom at scales $\Lambda$.
All particles constituting the SM have been found,
however, the symmetry-breaking sector is not yet fixed,
although the recent discovery at the LHC is very likely a Higgs particle \cite{Aad:2012tfa,Chatrchyan:2012ufa}. 

In the present paper we shall take this into account
by assuming a two-Higgs doublet model of type II (2HDM-II),
which allows for easy contact between supersymmetric models (SUSY),
and the SM with a single Higgs doublet
and also to heavy Higgs models
using nonlinear representations~\cite{Gunion:1989we,Donoghue:1978cj,Hall:1981bc}.
Focussing on quarks only and using the notation
\begin{eqnarray}
Q_L &=& \left( \begin{pmatrix}
               u_L \\
               d_L
               \end{pmatrix} ,
               \begin{pmatrix}
               c_L \\
               s_L
               \end{pmatrix} ,
               \begin{pmatrix}
               t_L \\
               b_L
               \end{pmatrix} \right) ,  \\ 
u_R &=& \left( u_R , c_R , t_R \right) , d_R = \left( d_R , s_R , b_R \right)  , 
\end{eqnarray}
the Yukawa couplings of the 2HDM-II model can be written
in terms of two Higgs doublet fields $\Phi_1$ and $\Phi_2$
with hypercharge $Y = 1$ and otherwise identical quantum numbers, as
\begin{equation} \label{Yuk} 
  -\lag^\thdm_{\rm Yuk}=\bar{Q}_L Y_D \Phi_1 d_R + \bar{Q}_L Y_U \tilde{\Phi}_2 u_R + {\rm H.c.}  , 
\end{equation}
where $\tilde\Phi$ is the charge-conjugated Higgs doublet
given by
\begin{equation}
  \tilde\Phi = i \tau_2 \Phi^\ast \ .
\end{equation}
``New physics'' beyond this SM-like dimension-4 piece is parametrized
in terms of higher-dimensional operators
\cite{Burges:1983zg,Leung:1984ni,Buchmuller:1985jz}
\begin{equation}
  \lag = \lag_{4D} + \frac{1}{\Lambda} \lag_{5D} + \frac{1}{\Lambda^2} \lag_{6D} + \ldots  , 
\end{equation}
where $\lag_{4D}\equiv\lag_{\rm SM}$,
and the new contributions
$\lag_{5D}$, $\lag_{6D}$, $\ldots$,
have to be symmetric under the SM gauge symmetry
$SU(3)_C \otimes SU(2)_L \otimes U(1)_Y$.
It turns out, that for quarks there is no dimension five operator compatible
with this symmetry
and thus the next-to-leading terms in the $\Lambda^{-1}$
expansion are of dimension six or higher.
The number of possible operators
is already quite large for a single  Higgs boson~\cite{Buchmuller:1985jz,Hansmann:2003jm}. 

We are going to consider this effective Lagrangian
at the top-quark mass scale,
which we identify with the electroweak scale $\mu \sim m_t$. 
Furthermore, we are interested in processes  leading to anomalous,
flavour-changing couplings of the top-quark to the gauge bosons,
and hence it is sufficient for us
to look at operators which are bilinear in the quark fields.
It is useful to classify the operators according to the helicities
of the quark fields, left-left (LL), 
right-right (RR) and left-right (LR).
Using this notation,
the Lagrangian can be written as
\begin{eqnarray}
  \lag = \lag_{\rm 2HDM}%
         &+& \frac{1}{\Lambda^2} \sum\limits_i \will{(i)}{\rm LL} \Op{(i)}{\rm LL} 
               + \frac{1}{\Lambda^2} \sum\limits_i \will{(i)}{\rm RR} \Op{(i)}{\rm RR}\notag\\
         &+& \frac{1}{\Lambda^2} \sum\limits_i \will{(i)}{\rm LR} \Op{(i)}{\rm LR} + \ldots  ,  \label{effth1}
\end{eqnarray}
where the $\will{(i)}{hh'}$ are generic coupling constants
which can in principle be calculated in specific models of new physics.   

In what follows we shall study  anomalous couplings
of flavour-changing currents involving top quarks to the gauge bosons.
To this end, it is sufficient to consider those dimension-6 operators
which are bilinear in the quark fields and which
-- after spontaneous symmetry breaking --
will induce vertices of the form $t \to q V$. 
However, not all of them are independent when applying the equations of motion. 
In particular, 
the operators involving three covariant derivatives
can be reduced either the ones with two derivatives or 
to four-fermion operators \cite{Bach:2012fb}. 
Furthermore, within the 2HDM-II model,
``flavour-changing neutral currents'' (FCNCs)
and large $CP$ violation are naturally suppressed by the imposed
discrete $\mathcal Z_2$ symmetry,
forbidding $\Phi_1\leftrightarrow \Phi_2$ transitions~\cite{Glashow:1976nt}.
Hence the set of independent operators for purely left-handed transitions 
can be chosen as~\cite{Buchmuller:1985jz}:
\begin{equation}
\begin{split}
\Op{ij(3)}{\rm LL} &=%
                  \left( \Phi_1^\dagger i D_\mu \Phi_1 \right)%
                     \left( \bar{Q}_{Li} \gamma^\mu Q_{Lj} \right) , \\
\Op{ij(4)}{\rm LL} &=%
                   \left( \Phi_2^\dagger i D_\mu \Phi_2 \right)%
                     \left( \bar{Q}_{Li} \gamma^\mu Q_{Lj} \right) , \\
\Op{ij(5)}{\rm LL} &=%
                   \left( \Phi_1^\dagger \tau_I i D_\mu \Phi_1 \right)%
                    \left( \bar{Q}_{Li} \tau_I \gamma^\mu Q_{Lj} \right) , \\
\Op{ij(6)}{\rm LL} &=%
                   \left( \Phi_2^\dagger \tau_I i D_\mu \Phi_2 \right)%
                     \left( \bar{Q}_{Li} \tau_I \gamma^\mu Q_{Lj} \right) ,
\end{split}
\end{equation}
and for purely right-handed transitions we have
\begin{equation}
 \begin{split}
\Op{ij(2)}{\rm RR}          &=%
                          \left( \Phi_2^\dagger i D_\mu \Phi_2 \right)%
                           \left( \bar{u}_{Ri} \gamma^\mu u_{Rj} \right) , \\ 
\Op{\prime  ij(2)}{\rm RR}  &=%
                          \left( \Phi_1^\dagger i D_\mu \Phi_1 \right)%
                           \left( \bar{d}_{Ri} \gamma^\mu d_{Rj} \right) , \\
\Op{ij(3)}{\rm RR}          &=%
                          \left( \tilde{\Phi}_2^\dagger i D_\mu \Phi_1 \right)%
                           \left( \bar{u}_{Ri} \gamma^\mu d_{Rj} \right) , 
\end{split}
\end{equation} 
where $D_\mu$ denotes the covariant derivative of the
SU(3)$_C$~$\otimes$~SU(2)$_L$~$\otimes$~U(1)$_Y$ gauge symmetry. 
For the transitions from left- to right-handed helicities we have
\begin{equation}
\begin{split}
\Op{ij(4)}{\rm LR}         &=%
                          \left( \bar{Q}_{Li} \sigma^{\mu\nu} \tau_I u_{Rj} \right)%
                           \tilde{\Phi}_2 W^I_{\mu\nu} + {\rm H.c.} ,  \\
\Op{ij(5)}{\rm LR}         &=%
                          \left( \bar{Q}_{Li} \sigma^{\mu\nu} u_{Rj} \right)%
                           \tilde{\Phi}_2 B_{\mu\nu} + {\rm H.c.} , \\ 
\Op{\prime  ij(4)}{\rm LR} &=%
                          \left( \bar{Q}_{Li} \sigma^{\mu\nu} \tau_I d_{Rj} \right)%
                           \Phi_1 W^I_{\mu\nu} + \rm{H.c.} , \\
\Op{\prime  ij(5)}{\rm LR} &=%
                          \left( \bar{Q}_{Li} \sigma^{\mu\nu} d_{Rj} \right)%
                           \Phi_1 B_{\mu\nu} + {\rm H.c.}  ,
\end{split}
\end{equation} 
where $W^ {I=1,2,3}_{\mu \nu}$ and
$B_{\mu \nu}$ denote the field strength of
SU(2)$_W$ and U(1)$_Y$ symmetries, respectively,
and $\sigma^{\mu\nu}=\frac{i}{2}[\gamma^\mu,\gamma^\nu]$.

In addition to these operators leading to anomalous weak couplings
we also can have anomalous coupling to gluons
which read
\begin{equation}
 \begin{split}
\ten{P}{ij(5)}{\rm LR}         &=%
                             \left( \bar{Q}_{Li} \sigma^{\mu\nu} T^a u_{Rj} \right)%
                              \tilde{\Phi}_2 G_{\mu\nu}^a + {\rm H.c.} , \\
\ten{P}{\prime  ij(5)}{\rm LR} &=%
                             \left( \bar{Q}_{Li} \sigma^{\mu\nu} T^a d_{Rj} \right)%
                              \Phi_1 G_{\mu\nu}^a + {\rm H.c.}  ,
\end{split}
\end{equation} 
where $T^a$ are the generators of SU(3)$_C$
and $G_{\mu \nu}^a$ is the gluon field strength.
Note that all these operators carry flavour indices $i,j$
and hence the coupling constants in Eq.~\eqref{effth1}
are actually $3 \times 3$ matrices in flavour space.
Thus it is evident that generic parametrization is pretty much useless
due to the large number of unknown parameters. 
\section{Minimal Flavour Violation}\label{sec:MFV} 
Data on flavour processes restricts the possible couplings
in Eq.~\eqref{effth1} severely.
Since currently there is no indication from flavour processes
of new effects at the TeV scale,
any NP at that scale must be ``minimally flavour violating''
\cite{Ali:1999we,Buras:2000dm},
i.e.\ the new physics couplings obey
the same flavour-suppression pattern as the standard model processes. 

The most economical way to implement this idea
has been advocated in Ref.~\cite{D'Ambrosio:2002ex},
where the flavour symmetry
\begin{equation} \label{FlavourGroup}
\mathcal G_F = \text{SU(3)}_{Q_L} \times \text{SU(3)}_{U_R} \times \text{SU(3)}_{D_R}  
\end{equation}
was introduced, which is
[up to -- for our purposes -- irrelevant U(1) factors]
the largest flavour symmetry
which is compatible with the SM.
Under this symmetry the quarks transform according to 
\begin{eqnarray}
Q_L  &\sim& (3,1,1)  ,  \\
u_R  &\sim& (1,3,1)  ,  \ %
d_R  \sim (1,1,3)  , 
\end{eqnarray}
while the SM gauge and the Higgs fields are singlets
with respect to $\mathcal G_F$.

In the SM, this symmetry is broken only by the Yukawa couplings 
$Y_U$ and $Y_D$ shown in Eq.~\eqref{Yuk}.
Following Ref.~\cite{D'Ambrosio:2002ex} these Yukawa couplings
can be introduced as spurion fields
with the transformation property 
\begin{equation} 
 Y_U  \sim (3,\bar{3},1)  ,  \ %
 Y_D  \sim (3,1,\bar{3})  , 
\end{equation}
such that the Yukawa interaction \eqref{Yuk} is rendered invariant.
``Freezing'' the spurion fields to the actual values of 
the Yukawa couplings yields the $\mathcal G_F$ symmetry breaking in the SM.

This spurion analysis can be extended to any new physics model
as well as to our effective field theory approach.  
To this end we insert the minimum number of spurions
into the set of higher-dimensional operators,
which are required to be Lorentz and gauge invariant
as well as flavour invariant. 
Thus
-- omitting some trivial structures which lead to flavour violation --
we get
\begin{widetext}
\begin{eqnarray}
\label{MNSpurionQQ} 
\sum_{i,j} \will{ij}{\rm LL} \left( \bar{Q}_{Li} \cdots Q_{Lj} \right) &=&  
    \bar{Q}_{L} \left[\alpha_{\rm LL} {\mathds{1}} +  \beta_{\rm LL} \, Y_U Y_U^\dagger
        + \eta_{\rm LL} \ Y_D Y_D^\dagger \right] 
         \cdots Q_{L}   ,   \\
\label{MNSpurionuu} 
\sum_{i,j} \will{ij}{\rm RR} \left( \bar{u}_{Ri} \cdots u_{Rj} \right) &=&
    \alpha_{\rm RR} \left( \bar{u}_{R}  \left[ Y_U^\dagger Y_D Y_D^\dagger Y_U \right]
         \cdots u_{R} \right) ,  \\ 
\label{MNSpuriondd} 
\sum_{i,j} \will{\prime  ij}{\rm RR} \left( \bar{d}_{Ri} \cdots d_{Rj} \right) &=&
    \beta_{\rm RR} \left( \bar{d}_{R}  \left[ Y_D^\dagger Y_U Y_U^\dagger  Y_D \right]
         \cdots d_{R} \right) ,  \\ 
\label{MNSpuriondu}
 \sum_{i,j} \will{\prime \prime ij}{\rm RR} \left( \bar{d}_{Ri} \cdots u_{Rj} \right) &=&
    \eta_{\rm RR} \left( \bar{d}_{R}  \left[ Y_D^\dagger  Y_U \right]
         \cdots u_{R} \right) ,  \\  
\label{MNSpurionQu} 
\sum_{i,j} \will{ij}{\rm LR} \left( \bar{Q}_{Li} \cdots u_{Rj} \right)  &=&
     \bar{Q}_{L} \left[ \lambda_U Y_U + \alpha_{\rm LR}  Y_D Y_D^\dagger Y_U \right]
         \cdots u_{R}   ,  \\ 
\label{MNSpurionQd} 
\sum_{i,j} \will{\prime  ij}{\rm LR} \left( \bar{Q}_{Li} \cdots d_{Rj} \right)  &=&
      \bar{Q}_{L} \left[  \lambda_D Y_D + \beta_{\rm LR} Y_U Y_U^\dagger Y_D \right]
          \cdots d_{R}   ,   
\end{eqnarray}
where the ellipses denote the Dirac, 
colour, and weak SU(2) matrices that appear in the operators.

The coefficients $\alpha_{\rm LL} \cdots \beta_{\rm LR}$
are expected to have a ``natural'' size.
The precise meaning of this statement depends on the way the NP effects enter the model.
A tree-level-induced NP effect
(e.g., a tree-level exchange of a new particle with mass $\Lambda$)
will induce coefficients $\alpha_{\rm LL} \cdots \beta_{\rm LR} \sim {\mathcal O}(1)$,
while loop-induced NP effects will suffer from the typical
loop-suppression factor $1/(16 \pi^2)$
and hence we would have $\alpha_{\rm LL} \cdots \beta_{\rm LR} \sim {\mathcal O}(10^{-2})$. 

The physical quark fields are the mass eigenstates,
which are defined in such a way that the neutral component 
of the terms proportional to $\lambda_U$ and $\lambda_D$
in \eqref{MNSpurionQu} and \eqref{MNSpurionQd} is diagonal,
since this contribution is exactly of the form of the mass terms in the SM Lagrangian.
This is achieved by picking a basis of the $Y_U$ and $Y_D$ where 
\begin{equation}
Y_U = Y_U^{\rm diag}   ,  \quad Y_D = V_{\rm CKM} Y_D^{\rm diag} , 
\end{equation} 
where $V_{\rm CKM}$ is the
Cabibbo-Kobayashi-Maskawa (CKM) rotation from the weak to the mass eigenbasis: 
$d_L^{\rm weak} = V_{\rm CKM} d_L^{\rm mass}$.

Resolving the terms in Eqs.~\eqref{MNSpurionQQ}--\eqref{MNSpurionQd}
into charged and neutral components,
one finds in terms of mass eigenstates for the charged component from Eq.~\eqref{MNSpurionQQ}
(from here all quark fields are mass eigenstates) 
\begin{equation}
\sum_{i,j} \will{ij}{\rm LL} \left( \bar{Q}_{Li} \cdots  \tau_+ Q_{Lj} \right) = %
           \bar{u}_L \cdots \left[ \alpha_{\rm LL} {\mathds 1}%
        +  \beta_{\rm LL}  \left(Y_U^{\rm diag} \right)^2%
        +  \eta_{\rm LL}   \left(Y_D^{\rm diag} \right)^2 \right]%
      V_{\rm CKM} d_L    , 
\end{equation} 
with $\tau_\pm = \frac12(\tau_1\pm i\tau_2)$,
while for the neutral components we get
\begin{eqnarray} 
   && \frac{1}{2} \sum_{i,j} \will{ij}{\rm LL} \left( \bar{Q}_{Li} \cdots (1+ \tau_3)  Q_{Lj} \right) =%
          \eta_{\rm LL} \bar{u}_L \cdots  V_{\rm CKM} \left( Y_D^{\rm diag} \right)^2 V_{\rm CKM}^\dagger  u_L  ,   \\ 
   && \frac{1}{2} \sum_{i,j} \will{ij}{\rm LL} \left( \bar{Q}_{Li} \cdots (1- \tau_3)  Q_{Lj} \right) =%
          \eta_{\rm LL} \bar{d}_L \cdots  V_{\rm CKM}^\dagger  \left( Y_U^{\rm diag} \right)^2 V_{\rm CKM}   d_L   . 
\end{eqnarray} 

For the remaining helicity combinations we get
\begin{eqnarray}
\sum_{i,j} \will{ij}{\rm RR} \left( \bar{u}_{Ri} \cdots u_{Rj} \right) &=&
    \alpha_{\rm RR} \left( \bar{u}_{R}  
    \left[ Y_U^{\rm diag}  V_{\rm CKM}  \left( Y_D^{\rm diag} \right)^2 V_{\rm CKM}^\dagger Y_U^{\rm diag}  \right]
         \cdots u_{R} \right) ,  \\ 
\sum_{i,j} \will{\prime  ij}{\rm RR} \left( \bar{d}_{Ri} \cdots d_{Rj} \right) &=&
    \beta_{\rm RR} \left( \bar{d}_{R}  
    \left[ Y_D^{\rm diag} V_{\rm CKM}^\dagger \left(Y_U^{\rm diag} \right)^2 V_{\rm CKM}  Y_D^{\rm diag} \right]
         \cdots d_{R} \right) ,   \\
 \sum_{i,j} \will{\prime \prime ij}{\rm RR} \left( \bar{d}_{Ri} \cdots u_{Rj} \right) &=&
    \eta_{\rm RR} \left( \bar{d}_{R}  \left[ Y_D^{\rm diag} V_{\rm CKM}^\dagger Y_U^{\rm diag} \right]
         \cdots u_{R} \right)  . 
\end{eqnarray}
Finally, Eqs.~\eqref{MNSpurionQu} and \eqref{MNSpurionQd}
have to be split into charged and neutral components. 
Omitting flavour diagonal contributions, we get 
\begin{eqnarray}
\left. \sum_{i,j} \will{ij}{\rm LR} \left( \bar{Q}_{Li} \cdots u_{Rj} \right) \right|_{\rm charged}   &=&
     \bar{d}_{L} \left[ \lambda_U V_{\rm CKM}^\dagger Y_U^{\rm diag}  
                + \alpha_{\rm LR}  \left(Y_D^{\rm diag} \right)^2 V_{\rm CKM}^\dagger Y_U^{\rm diag}  \right]
         \cdots u_{R}   ,  \\ 
\left. \sum_{i,j} \will{ij}{\rm LR} \left( \bar{Q}_{Li} \cdots u_{Rj} \right) \right|_{\rm neutral}   &=&
    \alpha_{\rm LR}   \bar{u}_{L} \left[  V_{\rm CKM} \left( Y_D^{\rm diag} \right)^2  V_{\rm CKM}^\dagger Y_U^{\rm diag} \right]
         \cdots u_{R}   ,  \\ 
\left. \sum_{i,j} \will{\prime  ij}{\rm LR} \left( \bar{Q}_{Li} \cdots d_{Rj} \right)  \right|_{\rm charged} &=&
      \bar{u}_{L} \left[  \lambda_D V_{\rm CKM}  Y_D^{\rm diag}  
           + \beta_{\rm LR} \left(Y_U^{\rm diag} \right)^2 V_{\rm CKM}  Y_D^{\rm diag}  \right]
          \cdots d_{R}   ,  \\
\left. \sum_{i,j} \will{\prime  ij}{\rm LR} \left( \bar{Q}_{Li} \cdots d_{Rj} \right)  \right|_{\rm neutral} &=& 
     \beta_{\rm LR}  \bar{d}_{L} \left[  V_{\rm CKM}^\dagger \left( Y_U^{\rm diag} \right)^2 V_{\rm CKM} Y_D^{\rm diag}  \right]
          \cdots d_{R}  . 
\end{eqnarray} 
\end{widetext}
Thus the flavour structure of the operators can be fixed by the assumption of MFV, 
and hence the number of independent couplings is reduced to
the number of operator structures listed in Sec.~\ref{sec:EA-2HD}.

Note that the entries in $Y_U^{\rm diag}$
and $Y_D^{\rm diag}$ are small except for the one entry in $Y_U^{\rm diag}$, 
corresponding to the top mass.
Also, for large $\tan \beta$, $Y_D^{\rm diag}$
may also contain a large entry related to the bottom mass.
It has been pointed out in Ref.~\cite{Feldmann:2008ja}
that this may spoil the expansion in powers of the spurion insertions.
We will not go into any details here
and restrict our analysis to the minimum number of spurion insertions.  

In most of the analyses using effective theory approaches,
unknown couplings are treated ``one at a time'',
which means that one coupling is varied with all other couplings set to zero.
In this paper we shall propose a slightly different scheme,
which automatically implements the relations among the couplings implied by MFV.
We will either set all the couplings to be unity,
$\alpha_{\rm LL} \cdots \beta_{\rm LR} \equiv 1$ (``tree-induced scenario''),
and vary the scale $\Lambda$,
or we will set $\alpha_{\rm LL} \cdots \beta_{\rm LR} \equiv 1/(16 \pi^2)$ (``loop-induced scenario''),
and vary the scale $\Lambda$.
Alternatively, we may fix the scale $\Lambda$ (e.g., at 1~TeV),
which means that we identify all the couplings $\alpha_{\rm LL} \cdots \beta_{\rm LR}$
and study the constraints on the remaining single parameter. 

We note that this scheme depends on the choice of the basis for the dimension-6 operators; however, 
the rationale behind this idea is that in a truly minimally flavour-violating scenario
all the remaning couplings should be natural, independently of the basis choice of the operators.
In turn this means that
-- up to the hierarchies implied by MFV --
no further hierarchical structures should emerge.
Without going into the details of a specific NP model,
there is no way to infer the detailed couplings;
if we want to stick to a model-independent approach
there is no alternative to such a crude scheme.  
\section{Decay Rates of Top Quarks}\label{sec:DRTop} 
In the remainder of the paper we shall focus on processes
with top quarks and their anomalous couplings to gauge bosons.
As mentioned in the previous section we use an effective theory approach
to study anomalous, flavour-changing top couplings at the weak scale $\mu \sim m_t$.
The MFV hypothesis allows us to predict relative sizes of
couplings for different flavours in the final state;
in turn, this may be used as a test of MFV in top decays,
once anomalous decays have been discovered.  
\subsection{Charged Currents}\label{sec:CC}
The first class of decays are the charged currents from couplings
of the form  $Wtq$, $q\in\{d,s,b\}$. 
Taking into account the various helicity combinations,
the effective interaction for the charged-current couplings has the general form
\begin{eqnarray}
\lag_{\rm eff} &=& \sum_{q = d,s,b}
                  \frac{g_2}{\sqrt2}
                 \biggl\{- \bar{q} \gamma^\mu \left( \coupL{q}{1} P_L + \coupR{q}{1} P_R \right) t W_\mu^- \notag\allowdisplaybreaks \\
                &&\!\! - (i \partial)_\nu \left[ \bar{q} \frac{i\sigma^{\mu\nu}}{M_W}
                     \left( \coupL{q}{2} P_L + \coupR{q}{2} P_R \right) t \right]  W_\mu^- \biggr\} , \label{eq:SMtqW}
\end{eqnarray}
where $P_{R/L}=\frac12 (1\pm\gamma_5)$
denote the chiral projectors.
Aplying the MFV hypothesis we get for the couplings 
\begin{equation}\label{eq:cWtb}
\begin{split}
\coupL{q}{1} =&   V_{tq}^\ast \left[%
                    1 + \frac{\alpha^{(5)}_{\rm LL}}{2}\frac{v_1^2}{\Lambda^2}%
	              + \frac{\alpha^{(6)}_{\rm LL}}{2}\frac{v_2^2}{\Lambda^2} \right]
	     =   V_{tq}^\ast + \delta \coupL{q}{1}  ,  \\
\coupR{q}{1} =&   V_{tq}^\ast  \eta^{(3)}_{\rm RR} \frac{m_q m_t}{\Lambda^2} ,  \\
\coupL{q}{2} =& 2 V_{tq}^\ast  \lambda_D^\ast  \frac{m_q v}{\Lambda^2} ,  \\
\coupR{q}{2} =& 2 V_{tq}^\ast  \lambda_U       \frac{m_t v}{\Lambda^2}   , 
\end{split}
\end{equation}
where $v_1 \equiv v \cos\beta$ and
$v_2 \equiv v \sin\beta$ are the vacuum expectation values
of the Higgs fields $\Phi_1$ and $\Phi_2$, respectively.
Note that we have kept the SM contribution in $L_1^q$
and defined $\delta\coupL{q}{1}$ to be the possible NP piece. 
Furthermore, the parameter $v^2 =v_1^2+v_2^2$
is fixed by the $W$-boson mass,
$\mws = \frac14g^2_2v^2$,
where $g_2$ is the weak SU(2) coupling constant,
or equivalently by the Fermi constant,
$G_F = 1/(\sqrt2 v^2)$.

As discussed above, the remaining unknown couplings in 
Eqs.~\eqref{eq:cWtb} are generically of order unity,
and hence in MFV we have the order-of-magnitude estimate 
\begin{equation} \label{MFV}
\begin{split}
\coupR{q}{1} &\sim \frac{2 m_q m_t}{v^2} \delta \coupL{q}{1}  ,  \
\coupL{q}{2}  \sim \frac{4 m_q}{v} \delta\coupL{q}{1}  , \\
\coupR{q}{2}                                                                                                                                                                                                          &\sim \frac{4 m_t}{v} \delta\coupL{q}{1}\ .
\end{split}
\end{equation} 
 
Although the renormalization-group flow
is not expected to change the orders of magnitude,
it is still important to know the renormalization effects for a quantitative analysis.  
We expect the relations~\eqref{MFV}
to hold at the high scale $\Lambda \gg \mu$,
and so we have to scale down to the scale of the top mass,
$\mu \sim m_t$,
where the measurement takes place. 

\begin{figure}
 \subfigure[\label{fig:ADim-1} self energies]{%
  \includegraphics[scale=.45]{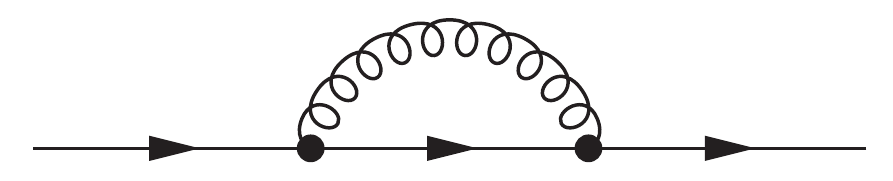}}
 \subfigure[\label{fig:ADim-2} vertex corrections]{%
  \includegraphics[scale=.45]{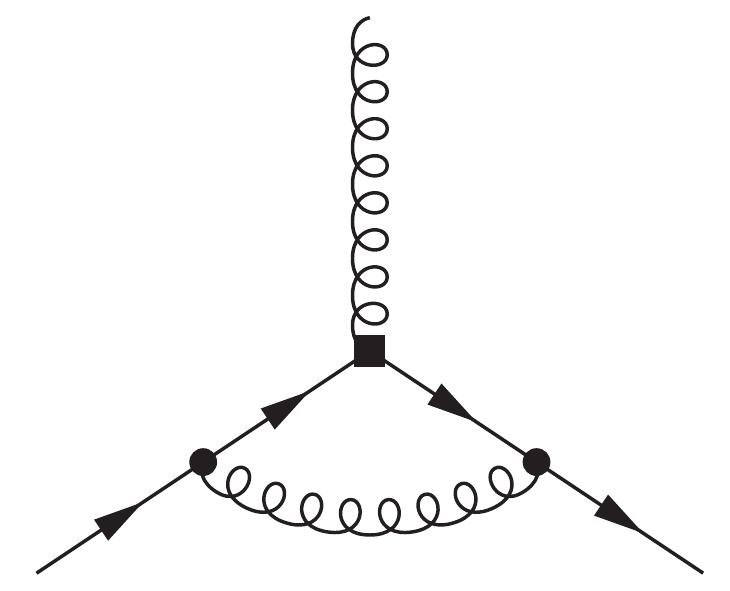}}
\caption{\label{fig:FD-ADim} Feynman diagrams for the
calculation of the anomalous dimension for
the charged electroweak case.}
\end{figure}
We focus on QCD effects only
and consider the diagrams shown in Figs.~\ref{fig:ADim-1} and \ref{fig:ADim-2}.
The left- and right-handed current do not have an anomalous dimension, and hence 
\begin{equation}
      \coupL{q}{1} (m_t) = \coupL{q}{1} (\Lambda) , %
\quad \coupR{q}{1} (m_t) = \coupR{q}{1} (\Lambda)\ . 
\end{equation} 

The helicity-changing contributions have an anomalous dimension
which has to be equal for both helicity combinations. 
To leading order one finds 
\begin{equation}
  \gamma^T(\alpha_s) = \frac{2 \alpha_s}{3 \pi} , 
\end{equation} 
which  yields for the running from
$\Lambda$ to $\mu \sim m_t$ for the two remaining couplings
\begin{equation}
  \begin{split}
  \coupL{q}{2}(m_t) &= \coupL{q}{2}(\Lambda) \left( \frac{\alpha_s(\Lambda)}{\alpha_s(m_t)} \right)^{\frac{4}{3 \beta_0}}  ,  \\
  \coupR{q}{2}(m_t) &= \coupR{q}{2}(\Lambda) \left( \frac{\alpha_s(\Lambda)}{\alpha_s(m_t)} \right)^{\frac{4}{3 \beta_0}}  , 
  \end{split}
\end{equation}
with 
\begin{equation}
\beta_0 = \frac{11 n_c - 2 n_f}{3} ,  
\end{equation} 
where $n_c$ and $n_f$ are the numbers of
colours and quark flavours, respectively.

From the effective operators we may calculate the amplitudes for top decays,
taking into account a possible NP effect. 
To this end, we use Eq.~\eqref{eq:SMtqW}
and compute the decay rates for the decay of an unpolarized top-quark
into a down-type quark $q$ and an on-shell $W$ boson.
The analysis of the $W$-decay products allows us to reconstruct its polarization, 
which is either longitudinal, left-, or right-handed.
The corresponding rates read \cite{Kane:1991bg,AguilarSaavedra:2006fy}
\begin{widetext}
\begin{eqnarray}
\Gamma (t \to q W_0) &=&
          \frac{g_2^2 |\vec{q}|}{32 \pi} \left\{ 
              \frac{m_t^2}{M_W^2} \left[
                 |\coupL{q}{1} |^2
               + |\coupR{q}{1} |^2 \right] \left(1-x_W^2 - 2 x_q^2 - x_W^2 x_q^2 + x_q^4 \right) 
               - 4 x_q {\rm Re}\left\{\coupL{q}{1}\coupR{q\ast}{1}\right\}  
              \right. \notag \\ 
            && \qquad \qquad \, \,  
               + \left[
                 |\coupL{q}{2} |^2
               + |\coupR{q}{2} |^2 \right] \left(1-x_W^2 + x_q^2\right)
               - 4 x_q {\rm Re}\left\{\coupL{q}{2}\coupR{q\ast}{2}\right\} 
              \notag \\ 
            && \qquad     
               - 2 \frac{m_t}{M_W} {\rm Re}
                  \left(\coupL{q}{1}\coupR{q\ast}{2}
                      + \coupL{q}{2}\coupR{q\ast}{1}\right) \left(1-x_W^2 - x_q^2\right) 
               \notag \\ 
            && \qquad \left. 
               + 2 \frac{m_t}{M_W} x_q {\rm Re}\left\{\coupL{q}{1}\coupL{q\ast}{2} + \coupR{q}{2}\coupR{q\ast}{1}\right\}
                      \left(1+x_W^2 - x_q^2\right)  \right\} \allowdisplaybreaks\\ 
\Gamma (t \to q W_{L/R}) &=& \frac{g_2^2 |\vec{q}|}{32 \pi} \left\{ 
                \left[ |\coupL{q}{1} |^2
                     + |\coupR{q}{1} |^2 \right]\left(1-x_W^2 + x_q^2\right)
              - 4 x_q {\rm Re} \left\{\coupL{q}{1}\coupR{q\ast}{1} \right\}  
              \right. \notag \\ 
            && \qquad 
               + \frac{m_t^2}{M_W^2} \left[
                    |\coupL{q}{2} |^2
                  + |\coupR{q}{2} |^2 \right] \left(1-x_W^2 - 2 x_q^2 - x_W^2  x_q^2 + x_q^4\right)
                  - 4 x_q {\rm Re}\left\{\coupL{q}{2}\coupR{q\ast}{2}\right\}  
              \notag \\ 
            && \qquad 
               - 2 \frac{m_t}{M_W}  {\rm Re}\left\{\coupL{q}{1}\coupR{q\ast}{2} + \coupL{q}{2}\coupR{q\ast}{1}\right\}
                      \left(1-x_W^2 - x_q^2\right) 
               \notag \\ 
            && \qquad \left. 
                + 2 \frac{m_t}{M_W} x_q {\rm Re}\left\{\coupL{q}{1}\coupL{q\ast}{2} + \coupR{q}{2}\coupR{q\ast}{1}\right\}
                       \left(1+x_W^2 - x_q^2 \right) \right\}  
              \notag \allowdisplaybreaks\\ 
            &\pm& \frac{g_2^2 m_t }{64 \pi} \frac{m_t^2}{M_W^2}
                \left\{  - x_W^2 \left[   |\coupL{q}{1} |^2 
                                        - |\coupR{q}{1} |^2 \right] 
                               + \left[   |\coupL{q}{2} |^2
                                        + |\coupR{q}{2} |^2 \right] \left(1-x_q^2\right) \vphantom{\frac11}\right.  
             \notag \allowdisplaybreaks\\ 
            && \left. \vphantom{\frac11} \qquad \quad \qquad
                  + 2 x_W     {\rm Re}\left\{\coupL{q}{1}\coupR{q\ast}{2} - \coupL{q}{2}\coupR{q\ast}{1}\right\} 
                  + 2 x_W x_q {\rm Re}\left\{\coupL{q}{1}\coupL{q\ast}{2} - \coupR{q}{2}\coupR{q\ast}{1})\right\}\right\}   
              \notag \\ 
            && \qquad \quad \qquad \quad \qquad\times \left(1-2 x_W^2 - 2 x_q^2 + x_W^4 - 2 x_W^2 x_q^2 + x_q^4 \right) 
\end{eqnarray} 
\end{widetext}
where the upper sign holds for left-handed
and the lower sign for right-handed $W$ bosons.
Furthermore, $x_q\equiv m_q/m_t$, $x_W \equiv M_W/m_t$,
\begin{equation}
 |\vec p| = \frac{m_t}{2}\sqrt{\lambda(1,\xq^2,\xw^2)} , 
\end{equation}
and the K\"all\'en function $\lambda(a,b,c) \equiv (a-b-c)^2-4bc$.

The total rate $\Gamma (t \to q W)$ is given by the sum 
\begin{eqnarray}
\Gamma (t \to q W) &=& \Gamma (t \to q W_0) + \Gamma (t \to q W_L )\notag\\
                   &&+ \Gamma (t \to q W_R )  ,  
\end{eqnarray}  
and the corresponding observables are the helicity fractions 
$F_0 = \Gamma (t \to q W_0) / \Gamma (t \to q W) $, $F_L = \Gamma (t \to q W_L) / \Gamma (t \to q W) $ and 
$F_R = \Gamma (t \to q W_R) / \Gamma (t \to q W) $. 
Using the condition $F_0 + F_R + F_L \equiv 1$
we get for the normalised differential decay rate \cite{Kane:1991bg,AguilarSaavedra:2006fy},
\begin{eqnarray}
 \frac{1}{\Gamma}\frac{d\Gamma}{d\cos\theta^\ast} &=&
             \frac{3}{8} \left(1-\cos\theta^\ast\right)^2 F_L
           + \frac{3}{8} \left(1+\cos\theta^\ast\right)^2 F_R \notag \\
         &&+ \frac{3}{4} \sin^2\theta^\ast F_0  , 
\end{eqnarray}
with the helicity angle $\theta^\ast$,
defined as the angle between the charged lepton three-momentum
in the $W$-boson rest frame
and the $W$-boson momentum in the top-quark rest frame.

The latest experimental measurements from
ATLAS \cite{ATLAS-CONF-2011-122}
and CMS \cite{CMS-PAS-TOP-11-020}
are shown in Table~\ref{tab:helfrac}.

\begin{figure}[ht] 
\includegraphics[width=.48\textwidth]{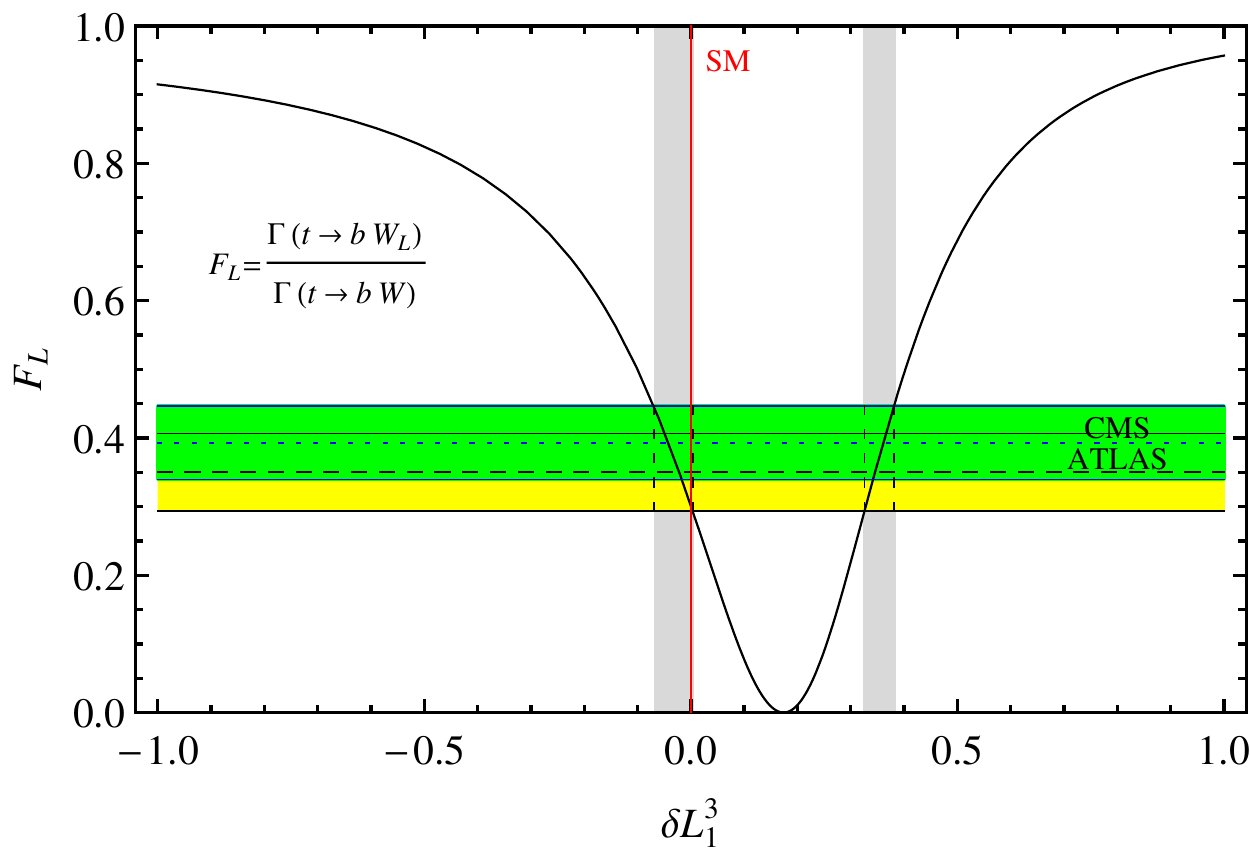} \includegraphics[width=.48\textwidth]{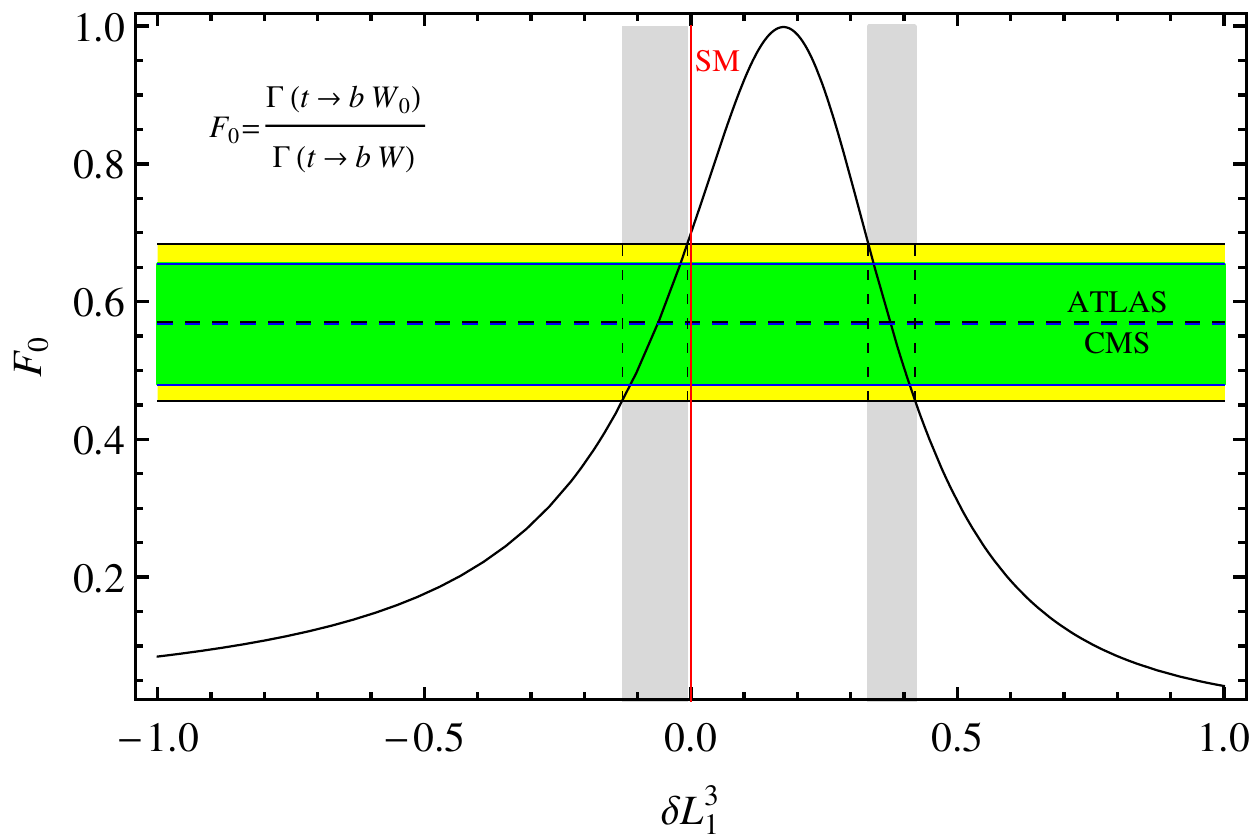}
\caption{%
Helicity fractions $F_L$ (top) and $F_0$ (bottom) for the decay $t \to b W$
as a function of a possible MFV new physics contribution with the coupling $\delta \coupL{q}{1}$.
The horizontal bands indicate the current data from the LHC experiments,
while the vertical bands indicate the currently allowed range for $\delta \coupL{q}{1} $.}
\label{Fig:tbW}
\end{figure}

\begin{table}[ht]
 \caption{\label{tab:helfrac} Helicity fractions @95\%~C.L.
from ATLAS and CMS for $t\to bW$ decay (see text for references).
The errors are statistical and systematic, respectively.}
\begin{ruledtabular}
\begin{tabular}{lll}
Fraction          & ATLAS                 & CMS\\\hline
$F_0$             & $0.57\pm0.07\pm0.09$  & $0.567\pm0.074\pm0.047$\\
$F_L$             & $0.35\pm0.04\pm0.04$  & $0.393\pm0.045\pm0.029$\\
$F_R$             & $0.09\pm0.04\pm0.08$  & $0.040\pm0.035\pm0.044$
\end{tabular}
\end{ruledtabular}
\end{table}
For a quantiative analysis we adopt the scheme described above.
This means in particular, that we take the relations \eqref{MFV} as equalities, 
and that we analyse the data in terms of the single quantity 
\begin{equation}
\delta \coupL{q}{1} = \alpha \frac{v^2}{\Lambda^2}   ,  
\end{equation} 
where $\alpha$ would be unity in a tree-induced scenario,
while $\alpha= 1/(16 \pi^2)$ in a loop-induced scenario. 
In Fig.~\ref{Fig:tbW} we plot the helicity fraction  $F_L$ and $F_R$
for $t \to b W$ as as function of $\delta \coupL{3}{1}$.
The standard model value corresponds to $\delta \coupL{q}{1} \equiv 0$
up to very small radiative corrections.
The colored bands indicate the data shown in Table~\ref{tab:helfrac}.
From this we infer that the SM value is well compatible with the current data.
However, given the current uncertainties,
there is still some room for a nonvanishing $\delta \coupL{q}{1}$. 
It is interesting to note that for both helicity fractions
a region around $\delta \coupL{q}{1} \sim 0.4$ is still allowed,
while the other region constrains 
$|\delta \coupL{q}{1}| \le  0.1$. 
 
The parameter $\delta \coupL{q}{1} $ still contains the dependence on $\Lambda$,
the scale of new physics.
Assuming a  value of $\Lambda \sim 1$~TeV 
we end up with $v^2 / \Lambda^2 \sim 0.1$.
This implies for NP scales around 1~TeV that the couplings in
Eq.~\eqref{eq:cWtb} can still be as large as unity,
implying that the current sensitivity cannot rule out MFV new physics effects at tree level.
In turn, in the loop-induced scenario there is still plenty of room for NP effects. 
\subsection{Neutral Currents}\label{sec:NC}
The study of FCNCs
such as $t\to qV$, $q\in\{u,c\}$, $V\in\{Z,\gamma, g\}$ 
is important in the context of NP analyses,
since the contribution of the SM is highly
Glashow-Iliopoulos-Maiani (GIM)
suppressed \cite{Glashow:1970gm}.
A measurement of such a process at the current level of sensitivity 
would clearly indicate new physics, in particular also implying non-MFV effects; 
the SM branching ratios ($\brf$) are in the region
$\brf(t\to qZ)\sim \mathcal O(10^{-13})$,
$\brf(t\to q\gamma)\sim \mathcal O(10^{-13})$ and
$\brf(t\to qg)\sim \mathcal O(10^{-11})$ \cite{Grzadkowski:1990sm}.
\begin{table}[ht]
 \caption{\label{tab:tqVOp}List of operators for $t \to qV$,
$q \in\{ u, c \}$, $V\in\{Z,\gamma,g\}$
transitions.}
\begin{ruledtabular}
\begin{tabular}{ll}
Decay	& Operator\\\hline
$t\to qZ$	& $\Op{ij(3)}{\rm LL}$, $\Op{ij(4)}{\rm LL}$, $\Op{ij(5)}{\rm LL}$, $\Op{ij(6)}{\rm LL}$\\ 
		& $\Op{ij(2)}{\rm RR}$, $\Op{ij(4)}{\rm LR}$, $\Op{ij(5)}{\rm LR}$\\\hline
$t\to q\gamma$	& $\Op{ij(4)}{\rm LR}$, $\Op{ij(5)}{\rm LR}$\\\hline
$t\to qg$	& $\ten{P}{ij(5)}{\rm LR}$
\end{tabular}
\end{ruledtabular}
\end{table}

The operators contributing to the FCNC interactions are listed in Table~\ref{tab:tqVOp}.
The experimental signatures of the various channels are quite different,
so we study the different processes separately in the following. 
\subsubsection{$t\to q Z$}\label{sec:tqZ}
The effective Lagrangian for the neutral currents
involving the $Z_0$
can be written as
\begin{eqnarray}
\lag_{\rm eff} &=& 
\frac{g_2}{\cos \theta_W} Z_\mu\notag\\
    &&\times \biggl\{  \bar{q} \gamma^\mu \left( \coupL{\prime  q}{1} P_L + \coupR{\prime  q}{1} P_R \right)t \notag\\
    &&	- \frac{(i\partial_\nu)}{M_Z}\left[ \bar{q}i \sigma^{\mu\nu} \left( \coupL{\prime  q}{2} P_L
	+ \coupR{\prime  q}{2} P_R \right) t \right] \biggr\} , \label{eq:tqZ}
\end{eqnarray}
with the couplings
\begin{widetext}
\begin{eqnarray}
\coupL{\prime  q}{1} &=&
             V_{qb} \ten{V}{\ast}{tb} \left[ 
                   \eta^{(3)}_{\rm LL} \frac{m_b^2}{\Lambda^2}
                 + \eta^{(4)}_{\rm LL}\frac{m_b^2}{\Lambda^2} \tan^2 \beta  \right. 
   \left. - \frac{ \eta^{(5)}_{\rm LL}}{2}\frac{m_b^2}{\Lambda^2}
          - \frac{ \eta^{(6)}_{\rm LL}}{2}\frac{m_b^2}{\Lambda^2} \tan^2 \beta \right]    ,  \allowdisplaybreaks\\
\coupR{\prime q}{1}  &=&
             V_{qb} \ten{V}{\ast}{tb}
               \alpha^{(2)}_{\rm RR} \frac{m_b^2}{\Lambda^2} \frac{m_q m_t }{v^2} \frac{1}{\sin^2 \beta} , \allowdisplaybreaks \\
\coupL{\prime q}{2} &=&
           2 V_{qb} \ten{V}{\ast}{tb}
              \frac{m_q}{v} \frac{1}{\sin^2 \beta}  
                 \biggl( \cos \theta_W \alpha^{(4)\ast}_{\rm LR} \frac{m_b^2}{\Lambda^2} 
                       - \sin \theta_W \alpha^{(5)\ast}_{\rm LR} \frac{m_b^2}{\Lambda^2} \biggr) ,  \allowdisplaybreaks\\
\coupR{\prime q}{2} &=&
           2 V_{qb} \ten{V}{\ast}{tb}
              \frac{m_t}{v} \frac{1}{\sin^2 \beta} 
                 \biggl( \cos \theta_W \alpha^{(4)}_{\rm LR}\frac{m_b^2}{\Lambda^2}
                       - \sin \theta_W \alpha^{(5)}_{\rm LR}\frac{m_b^2}{\Lambda^2} \biggr) , 
\end{eqnarray}
\end{widetext}
where $\theta_W$ is the Weinberg angle. 

As an order-of-magnitude estimate,
it is worthwhile to note that MFV leads to a significant GIM-like suppression
of this coupling by a factor of $m_b^2 / \Lambda^2$,
which is much smaller than the ``natural'' value of the coupling $v^2 / \Lambda^2$.
Furthermore, also for the relative sizes of the couplings for the various helicity combinations,
we get an MFV prediction $\tan \beta \sim 1$ or larger
\begin{equation}\label{eq:ttoqZest}
\begin{split}
   \coupR{\prime q}{1}	&\sim
                \coupL{\prime q}{1} \frac{m_q m_t }{v^2} \frac{1}{\tan^2 \beta}  , \\
   \coupL{\prime q}{2}	&\sim
                \coupL{\prime q}{1} \frac{m_q}{v} \frac{1}{\tan^2 \beta} ,  \\     
   \coupR{\prime q}{2}	&\sim
                \coupL{\prime q}{1} \frac{m_t}{v} \frac{1}{\tan^2 \beta} ,  
\end{split}
\end{equation} 
Finally we note that there is also a loop-induced contribution from the SM
which has been calculated in Ref.~\cite{AguilarSaavedra:2002ns}.
However, the relevant vertex cannot be expressed as a local operator,
and hence the expressions are quite cumbersome.
On the other hand, since the SM is by construction MFV,
the same suppression factors as for the new physics contribution will appear,
and since the SM rates are tiny compared to the current experimental limit
we take for the SM contribution the simple estimates  
\begin{eqnarray} \label{SMsimple}
    \coupL{\prime  q}{1 {\rm SM}} &=&
            V_{qb} \ten{V}{\ast}{tb}  \frac{1}{16 \pi^2}  \frac{m_b^2}{v^2} ,  \\ 
    \coupR{\prime  q}{1 {\rm SM}} &=&
            V_{qb} \ten{V}{\ast}{tb}  \frac{1}{16 \pi^2}  \frac{m_b^2}{v^2} \frac{m_q m_t }{v^2} ,  \notag  \\ 
    \coupL{\prime  q}{2 {\rm SM}} &=&
    \coupR{\prime  q}{2 {\rm SM}}  =
            V_{qb} \ten{V}{\ast}{tb}  \frac{1}{16 \pi^2}  \frac{m_b^2}{v^2} \frac{m_t }{v}  , \notag
\end{eqnarray} 
which is numerically very close to the real calculation,
i.e.\ it yields a branching fraction of the order 
\begin{equation}
\brf_{\rm SM} (t \to c Z)
      \sim \left| \frac{1}{16 \pi^2} V_{cb}  \frac{m_b^2}{v^2} \right|^2 \sim 6 \times 10^{-15} \ .
\end{equation}  

\begin{figure}[ht] 
\includegraphics[width=.48\textwidth]{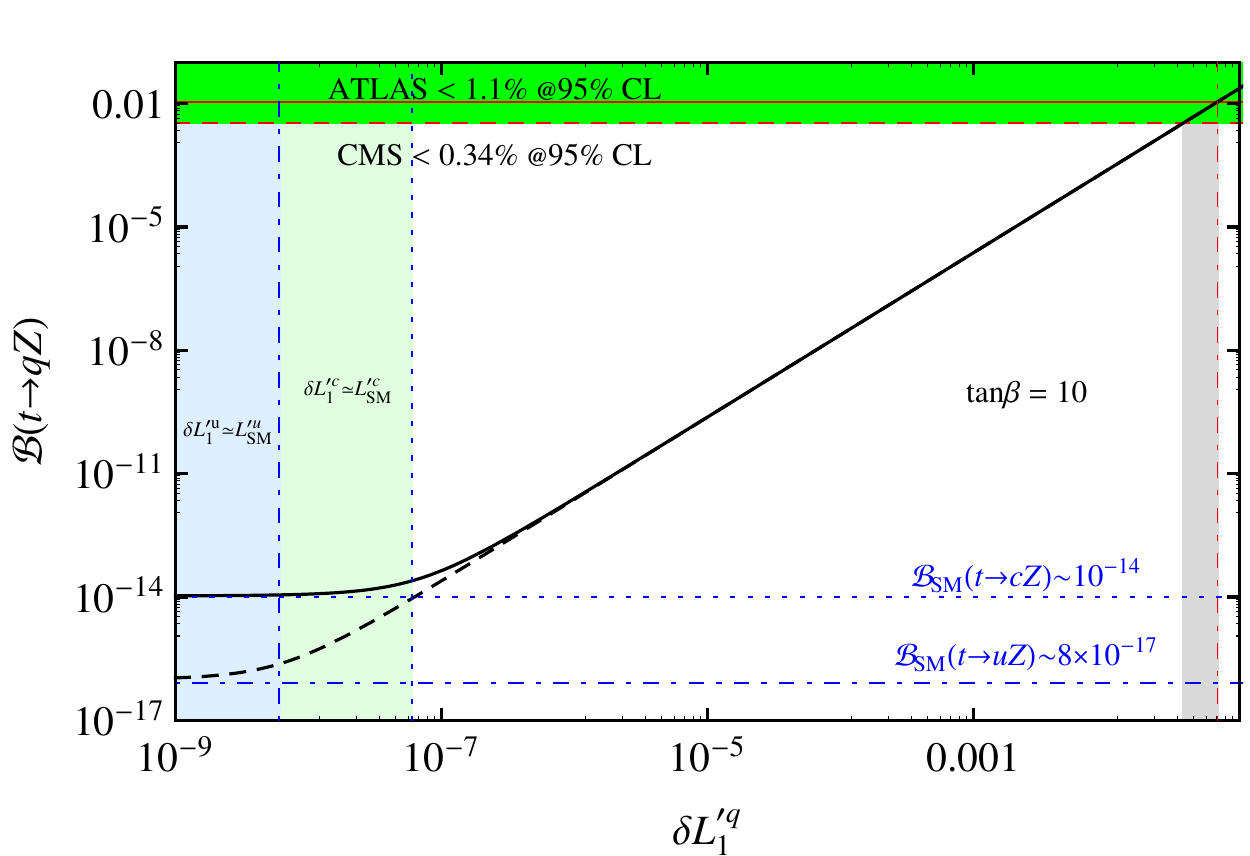}  
\caption{%
The branching fraction for $t \to Z q$
as a function of the coupling $\coupL{\prime  q}{1}$
and $\tan\beta = 10$. 
The horizontal lines indicate the expectation within the SM
and the current limits imposed by
ATLAS \cite{CortesGonzalez:2012ym,ATLAS-CONF-2011-154}
and CMS \cite{CMS-PAS-TOP-11-028}.
}
\label{fig:t2cZ}
\end{figure}

In order to perform a numerical analysis
we proceed similarly to the case of charged currents.
We take the approximate relations \eqref{eq:ttoqZest} 
as exact equations and express everything in terms of the new physics coupling $\coupL{\prime q}{1}$. 
Including the standard model estimate on the basis 
of the naive estimate \eqref{SMsimple} we show in Fig.~\ref{fig:t2cZ}
the branching fraction for $t \to q Z$, $q\in\{u,c\}$. 

We note that the natural size of  $\coupL{\prime q}{1}$ is in MFV
given by $V_{qb} V_{tb}^\ast m_b^2 / \Lambda^2 $,
assuming tree-level FCNC effects. 
This  is for $\Lambda \sim 1$~TeV about $10^{-6}$,
which is one order of magnitude above the SM value.
Loop-induced new physics effects would show up 
in MFV for $\Lambda \sim 1$~TeV only at a level of
$\coupL{\prime q}{1} \sim 10^{-9}$,
which is far below the SM value.
In turn, the current experimental limit 
implies $\coupL{\prime q}{1} \le 0.01$,
which is far above the prediction of any MFV scenario. 
\subsubsection{$t\to q\gamma$ and $t\to q g$}\label{sec:tqgamma}
The possible couplings for photonic transitions
are more restricted due to electromagnetic gauge invariance.  
Extracting the relevant terms form the dimension-6 operators we find
for the effective Lagrangian
\begin{eqnarray}
  \lag_{\rm eff} = -e  \mathcal A_\mu 
       \frac{(i \partial_\nu)}{m_t} \left[
             \bar{q} i \sigma^{\mu\nu} \left(
                   L^{(2)}_q P_L
                 + R^{(2)}_q P_R \right) t \right] , \label{eq:tqgamma}
\end{eqnarray}
where $e$ and $\mathcal A_\mu$ are the electron charge
and the electromagnetic field, respectively.

For the $\gamma tq$ vertex
we get only two independent anomalous coupling constants,
\begin{widetext}
\begin{equation}
 \begin{split}
  \coupL{(2)}{q} &=
           \frac{4 V_{qb} \ten{V}{\ast}{tb}}{e} \left(
                     \sin \theta_W \will{32(4)\ast}{\rm LR}
                   + \cos \theta_W \will{32(5)\ast}{\rm LR} \right)
                       \frac{m_b^2}{\Lambda^2}\frac{m_qm_t }{v_1^2}  ,  \\
  \coupR{(2)}{q} &=
           \frac{4 V_{qb} \ten{V}{\ast}{tb}}{e} \left(
                     \sin \theta_W \will{23(4)}{\rm LR}
                   + \cos \theta_W \will{23(5)}{\rm LR} \right)
                        \frac{m_b^2}{\Lambda^2 }\frac{m_t^2}{v_1^2}  .
\end{split}
\end{equation}
\end{widetext}
Clearly, in a MFV scenario, the right-handed top quark
yields the dominant contribution,
\begin{equation}\label{eq:tqgammaest}
    \coupL{(2)}{q} \sim \frac{m_q}{m_t} \coupR{(2)}{q}  \ .
\end{equation} 
Taking this as an equation,
we can express the rate and the branching fractions
in terms of the single coupling 
$\coupR{(2)}{q}$. 
\begin{figure}[ht] 
\includegraphics[width=.48\textwidth]{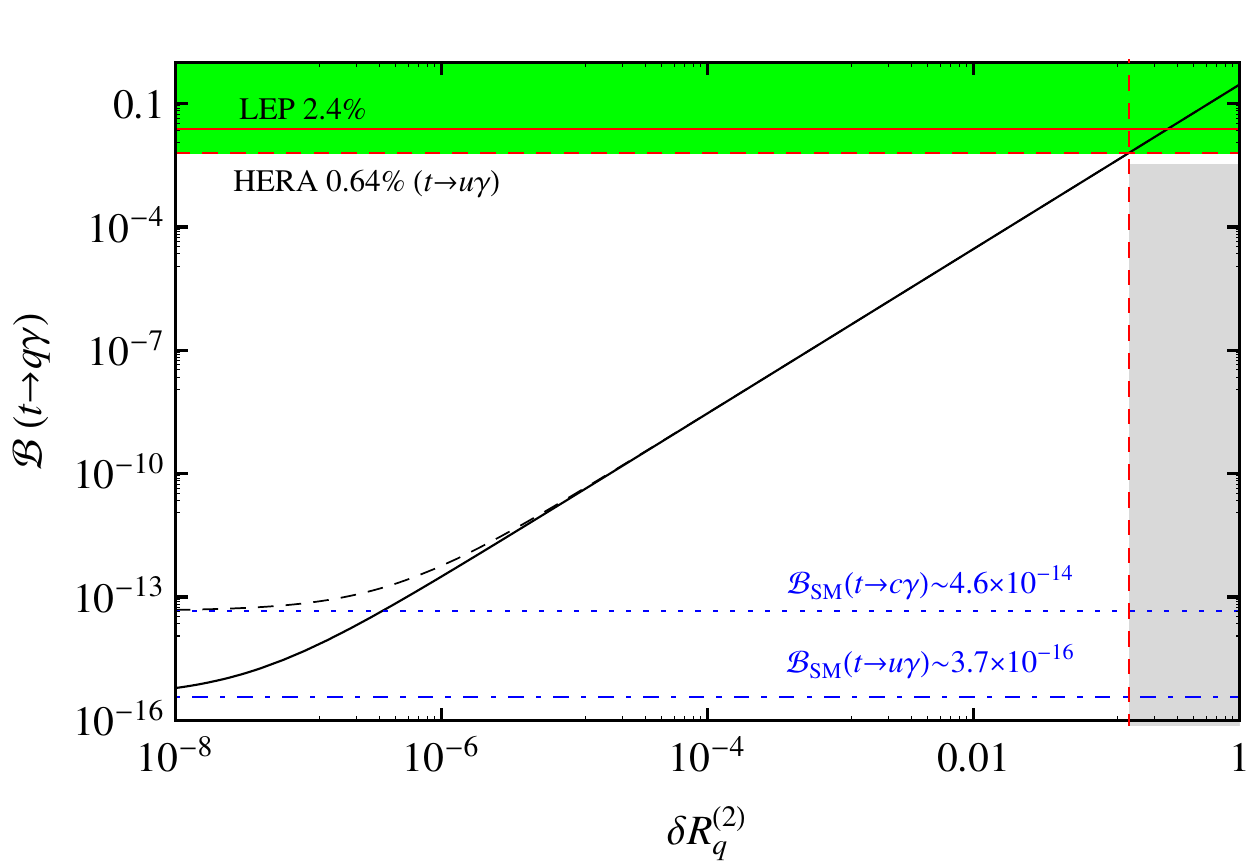}  
\caption{%
The branching fraction for $t \to q \gamma$
as a function of the coupling $\coupL{\prime  q}{1}$. 
The horizontal lines indicate the expectation within the SM
and the current limits imposed by
LEP \cite{Heister:2002xv,Abdallah:2003wf,Abbiendi:2001wk,Achard:2002vv,LEP-Exotica:WG-2001-01}
and HERA $(t\to u\gamma)$ \cite{Aaron:2009vv},
and also from 
Tevatron: 3.2\% \cite{Abe:1997fz}
which is not shown.
}
\label{fig:t2cgamma}
\end{figure} 

The standard model value for this process is very small.
This can be inferred already from a simple order-of-magnitude estimate
by just collecting the factors for the loop suppression and the CKM and mass factors. 
One obtains 
\begin{eqnarray}
\brf_{\rm SM} (t \to q \gamma)
       &\sim& \left| \frac{\alpha}{\pi} V_{tb} V_{qb}^\ast  \frac{m_b^2}{M_W^2} \frac{m_t^2}{v^2} \right|^2 \notag\\
       &\sim&       \begin{cases}
                     4 \times 10^{-14} & {\rm for} \quad t \to c \gamma \\
                     4 \times 10^{-16} & {\rm for} \quad t \to u \gamma
                    \end{cases}   
\end{eqnarray}  
which is numerically close to the values obtained from the full calculation. 

As in the case of $t \to Z q$, the MFV expectation is very small.
Assuming a tree-like scenario, $\coupR{(2)}{q}$
is of the order $V_{qb} \ten{V}{\ast}{tb} m_b^2 / \Lambda^2$.
For a new physics scale $\Lambda \sim 1$~TeV
we end up with a typical expectation
$\coupR{(2)}{c} \sim 10^{-7}$ and
$\coupR{(2)}{u} \sim 10^{-8}$
for the loop-induced scenario;
this is even smaller
by a factor of $1/(16 \pi^2)$.
Thus, if nature is minimally flavour violating,
but otherwise the couplings have natural sizes,
the current limits are several orders of magnitude above the expectations. 

The case $t \to g q$ is very similar to $t \to q \gamma$;
the only difference is the larger strong coupling
and larger QCD renormalization effects.
The one-loop QCD renormalization is given in Appendix~\ref{appsec:QCDRen};
although the effects can be sizable,
they are still far away from being relevant in the current experimental situation. 

Due to QCD gauge invariance the effective interaction
has a similar form as the photonic operator,
\begin{eqnarray}
  \lag_{\rm eff} &=& -g_s  G_\mu^a 
        \frac{1}{m_t} \notag\\
        && \times (-i \partial_\nu)
              \left[ \bar{q}  T^a i \sigma^{\mu\nu} 
                    \left( \tilde{L}^{(2)}_q P_L
                         + \tilde{R}^{(2)}_q P_R \right) t \right] , \label{eq:tqg}
\end{eqnarray}
with 
\begin{align}
  \ten{\tilde L}{(2)}{q} &= 
        V_{qb} V^\ast _{tb} 
             \frac{4}{g_s} \ten{K}{32(5)\ast}{\rm LR} \frac{m_b^2}{\Lambda^2}\frac{m_qm_t}{v_1^2} ,  \\
   \ten{\tilde R}{(2)}{q} &=
        V_{qb} V^\ast _{tb} 
             \frac{4}{g_s} \ten{K}{23(5)}{\rm LR} \frac{m_b^2}{\Lambda^2}\frac{m_t^2}{v_1^2} , 
\end{align}
where $\ten{G}{a}{\mu}$ is the field strength-tensor
of the gluon fields and $T^a$ are the Gell-Mann matrices. 

\begin{figure}[ht] 
\includegraphics[width=.48\textwidth]{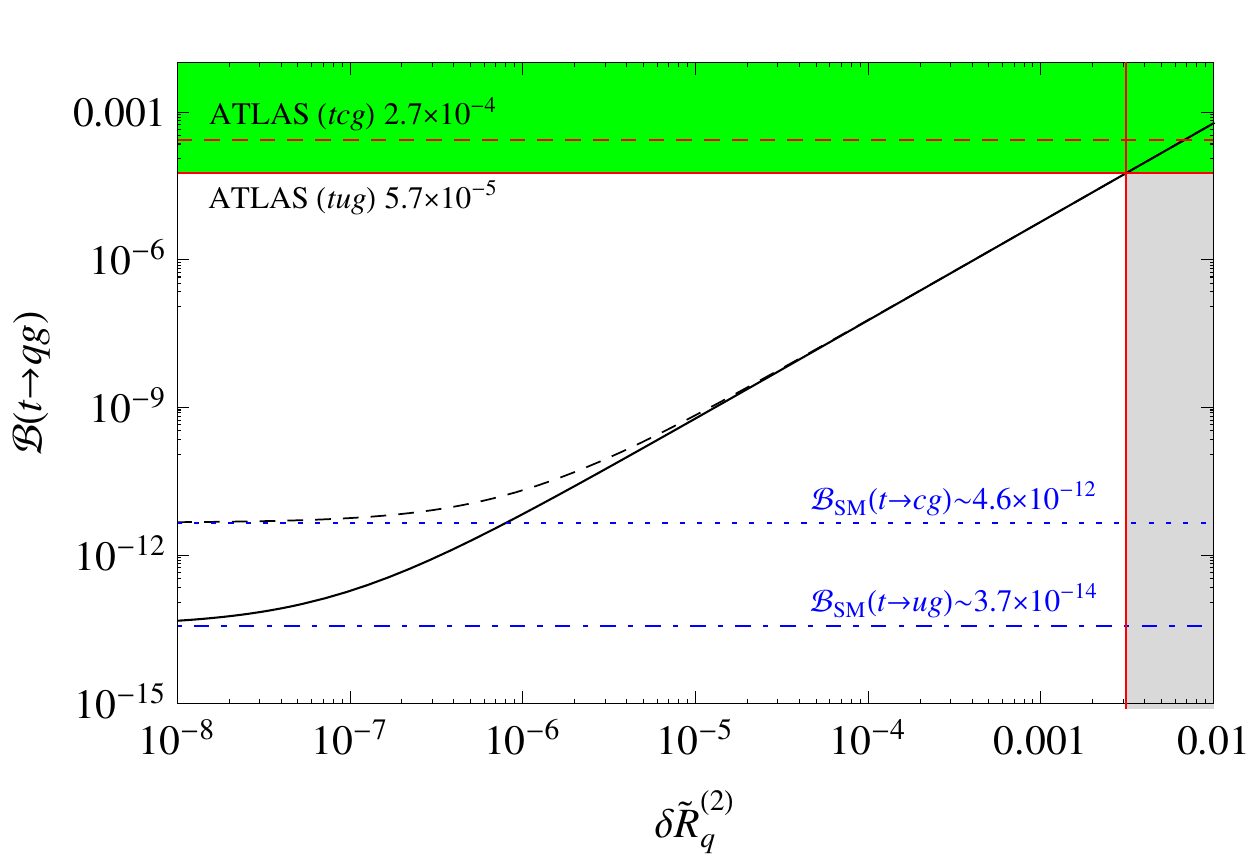}  
\caption{%
The branching fraction for $t \to q g$
as a function of the coupling $\tilde{R}^{(2)}_q$. 
The horizontal lines indicate the expectation
within the SM and the current limits imposed by
ATLAS \cite{Collaboration:2012gd}.
}
\label{fig:t2cg}
\end{figure}

For $\ten{K}{32(5)}{\rm LR}, \ten{K}{23(5)}{\rm LR} \sim 1$
we get the same order-of-magnitude relation for the couplings
as for the photonic case
\begin{equation}\label{eq:tqgest}
    \tilde{L}^{(2)}_q \sim \frac{m_q}{m_t} \tilde{R}^{(2)}_q  ,
\end{equation}
and we shall again use this as an equality
to perform an MFV analysis in terms of the single variable. 
Figure~\ref{fig:t2cg} shows the current status. 

The estimate in the standard model
is the same as for $t \to q \gamma$
with the electromagnetic coupling replaced by the strong one,
\begin{eqnarray}
\brf_{\rm SM} (t \to q g)
      &\sim& \left| \frac{\alpha_s}{\pi} V_{tb} V_{qb}^\ast  \frac{m_b^2}{M_W^2}  \frac{m_t^2}{v^2} \right|^2 \notag\\
      &\sim& \begin{cases}
               6 \times 10^{-12} & {\rm for} \quad t \to c g \\
               4 \times 10^{-14} & {\rm for} \quad t \to u g
             \end{cases} 
\end{eqnarray}  
which again is close to the result of the full calculation. 
Concerning the expectations of the MFV scenario,
one arrives at the same conclusions as for $t \to q \gamma$:
the MFV expectations are still several orders of magnitude away from the experimental sensitivity. 
\section{Comparison to other constraints}
Aside from direct searches, indirect constraints
may also be obtained from both electroweak as well as from $B$ and $K$ decays. 

The assumption of MFV also links the flavour physics of the top quark
with that of the bottom and the strange quark.
Overall, the constraints from $B$ and $K$ physics
which are discussed at length in Refs.~\cite{D'Ambrosio:2002ex,Hurth:2008jc},
are much more restrictive than the ones obtained from the current data on the top quark.
In turn, any effect that could be seen in top decays
at the current level of precision would indicate a deviation from MFV. 

A certain loophole in the MFV argument emerges due to the large top-quark mass,
implying an order unity Yukawa coupling for the top.
This may be closed by employing a nonlinear realization of MFV \cite{Feldmann:2008ja};
however, as discussed in Ref.~\cite{Kagan:2009bn}, some effects may become visible in the top 
sector for large values of $\tan \beta$.   

Another way of testing anomalous $t \to b W$ couplings is in loop-induced $B$ decays.
It has been shown in Ref.~\cite{Grzadkowski:2008mf} that in particular the decay $B \to X_s \gamma$
is quite sensitive to an anomalous $t \to b W$ coupling resulting in a limit.
The combined analyses performed in Refs.~\cite{Drobnak:2011aa,Drobnak:2011wj},
including also other FCNC processes of $B$ mesons, 
arrives at bounds for anomalous $t \to b W$ couplings
which are again stronger than what can be obtained from the current 
direct measurements at the LHC. 

Also the precision data from the electroweak sector constrain
possible nonstandard top couplings.
Such an analysis, using the precision data on the oblique parameters 
of the electroweak sector, has been performed in Ref.~\cite{Zhang:2012cd}.
The constraints obtained in this way are comparable to
the ones obtained from flavour decays and 
thus are again much stronger than the bounds from the currently available direct measurements. 
\section{Conclusions}\label{sec:Conc}
We have discussed anomalous,
flavour-changing top decays in a model-independent way
by employing an effective field theory approach.
The new element in our analysis is the implementation of minimal flavour violation 
by setting up a simple scheme with few parameters,
which obeys the MFV hierarchical structure.
We have calculated  the decay rates for the charged current
$t\to qW$, $q\in\{d,s,b\}$,
as well as for the FCNC couplings
$tqV$, $q\in\{u,c\}$, $V\in\{Z,\gamma,g\}$,
in terms of the effective couplings for different helicities
under the assumption that the top quark is produced
in a high-energy collision as a quasi-free particle.  
Comparing such a scenario with the present data shows that there is still plenty 
of room for NP effects in the anomalous, flavour-changing top couplings,
in particular for the flavour-changing neutral current decays. 
\section*{Acknowledgements}
One of us (S.F.) would like to thank M. Jung for
help-full discussions and comments.
This work was supported by
the German Ministry for Research and Education
(BMBF, Contract No. 05H12PSE).
\begin{widetext}
\appendix
\section{QCD Renormalization}\label{appsec:QCDRen}
\begin{figure}[ht]
\subfigure[\label{fig:FD-AD-tqVa}]{%
  \includegraphics[scale=.48]{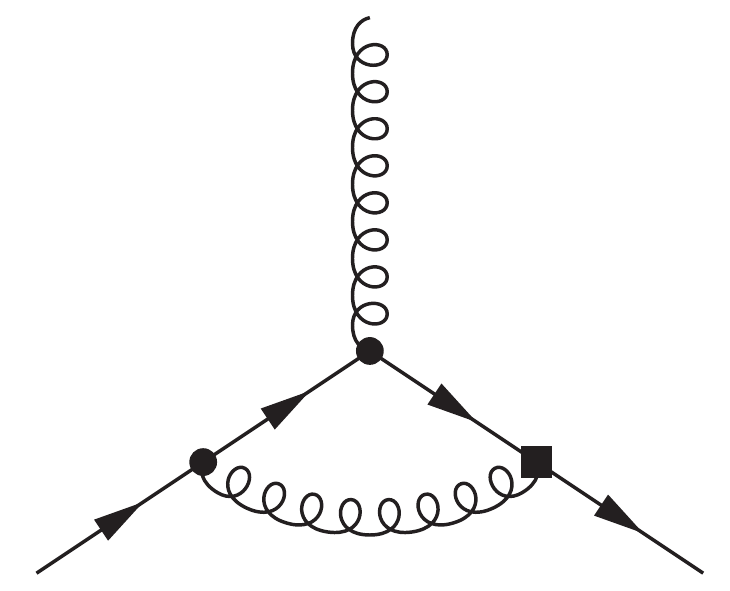}}
\subfigure[\label{fig:FD-AD-tqVb}]{%
  \includegraphics[scale=.48]{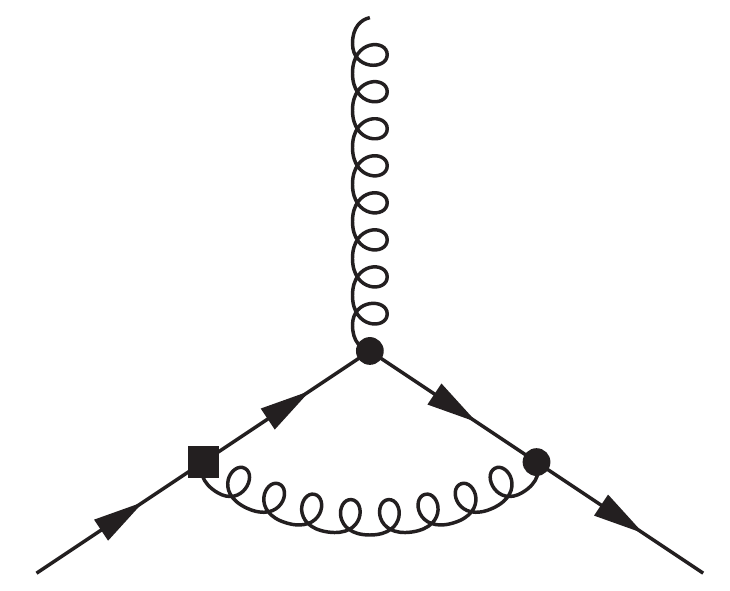}}
\subfigure[\label{fig:FD-AD-tqVc}]{%
  \includegraphics[scale=.48]{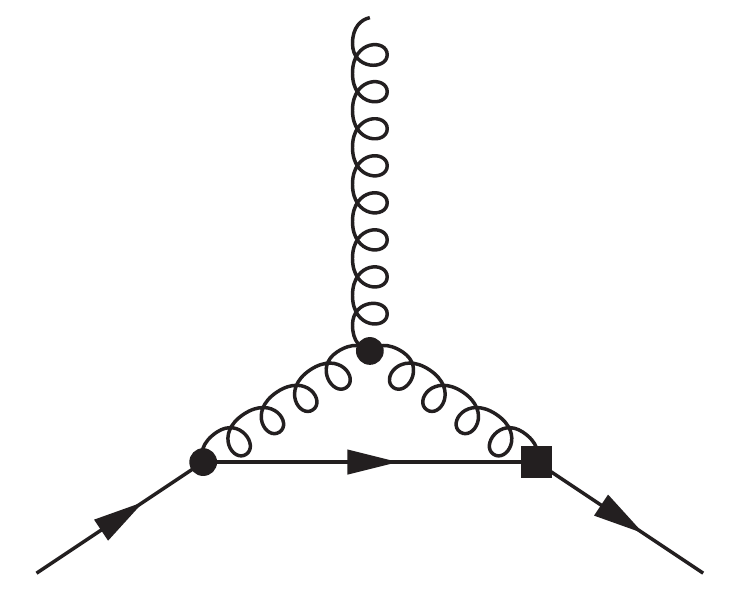}}
\subfigure[\label{fig:FD-AD-tqVd}]{%
  \includegraphics[scale=.48]{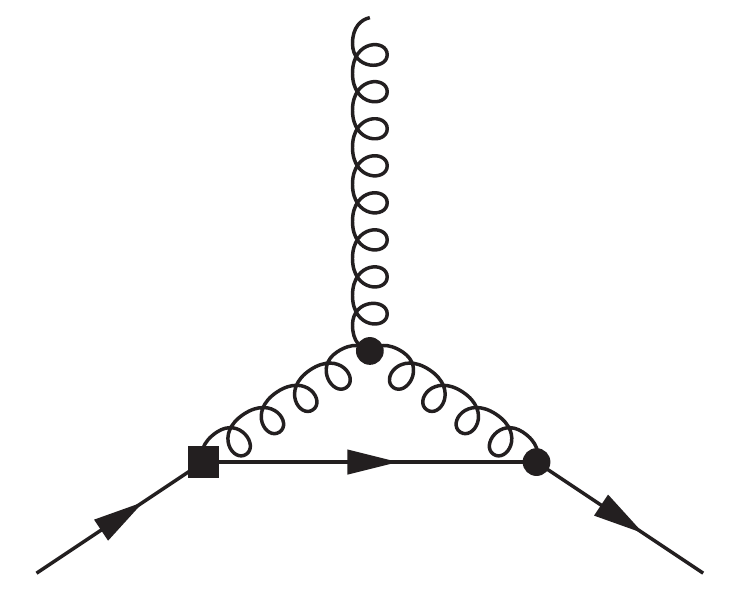}}
\subfigure[\label{fig:FD-AD-tqVe}]{%
  \includegraphics[scale=.48]{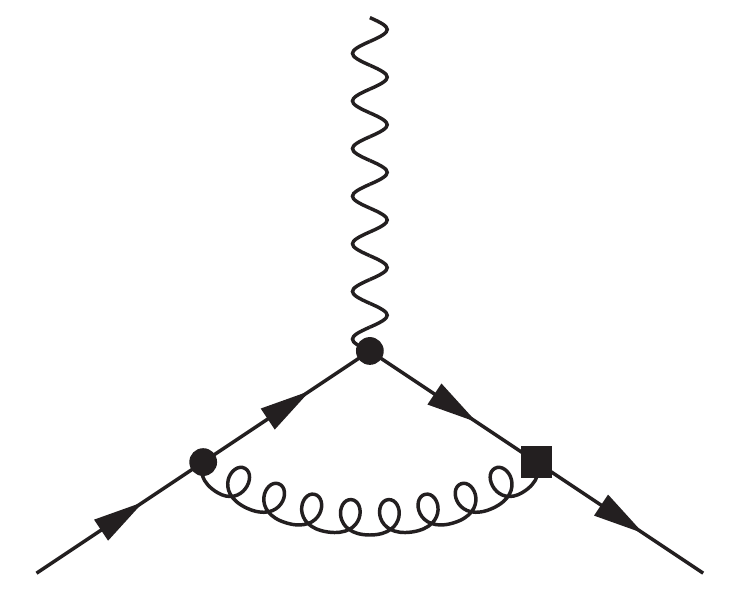}}
\subfigure[\label{fig:FD-AD-tqVf}]{%
  \includegraphics[scale=.48]{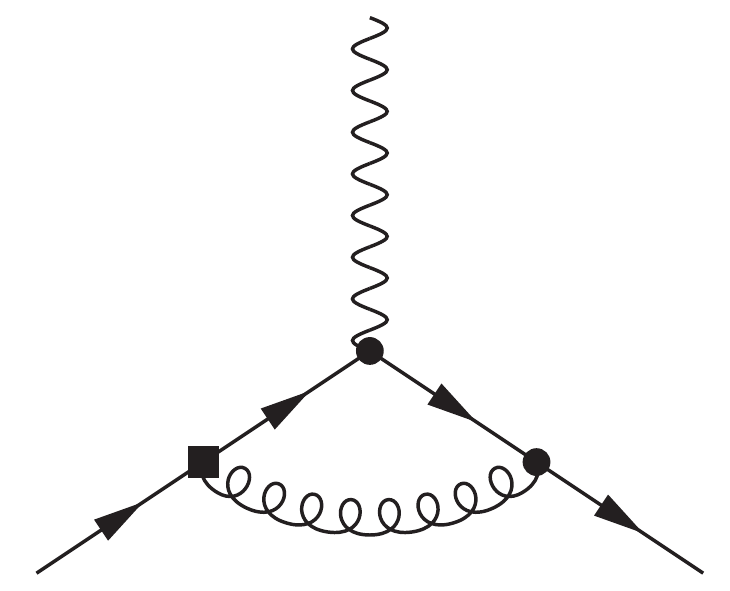}}
\caption{\label{fig:FD-AD-tqV}Feynman diagrams for the calculation
of the anomalous dimension for the strong decays.
Diagrams (e) and (f) include a mixing of strong
and electroweak interactions. }
\end{figure} 
The calculation of the anomalous dimension
for the $t\to qV$ transitions proceeds along the same lines
as we discussed in Sec.~\ref{sec:CC} for the charged current.
However, for the QCD-like structure of the generalised
$gtq$ vertex, Eq.~\eqref{eq:tqg},
additional diagrams have to be taken into account.
All Feynman diagrams shown in Fig.~\ref{fig:FD-AD-tqV}
-- where Figs.~\ref{fig:FD-AD-tqVe} and \ref{fig:FD-AD-tqVf}
represent the mixing of strong and electroweak operators --
have to be calculated. 
We define
\begin{eqnarray}
 \vec{\will{}{}}&=&\begin{pmatrix} 
                  \ten{\tilde L}{(2)}{q}\\
		  \ten{\tilde R}{(2)}{q}\\
		  \coupL{(2)}{q}\\
		  \coupR{(2)}{q}\\
                  \coupL{\prime q}{2}\\
                  \coupR{\prime q}{2}
                 \end{pmatrix}  ,  \allowdisplaybreaks\\
 \vec{\Op{}{}} &=& \left(-i\partial_\nu\right) \left( \bar q\sigma^{\mu\nu} \begin{pmatrix}
		   \frac{1}{m_t} T_a P_L \ten{G}{a}{\mu} \\
 		   \frac{1}{m_t} T_a P_R \ten{G}{a}{\mu} \\
 		   \frac{1}{m_t} P_L \mathcal A_\mu \\
  		   \frac{1}{m_t} P_R \mathcal A_\mu \\
		   \frac{1}{M_Z} P_L Z_\mu \\
 		   \frac{1}{M_z} P_R Z_\mu \\
		 \end{pmatrix} t \right) , 
\end{eqnarray}
and for the $6\times 6$
anomalous-dimension matrix
for the Wilson coefficients,
using Feynman gauge, we get
\begin{equation}
 \gamma^T (\mu) = \frac{2\alpha_s(\mu)}{3\pi}
     \begin{pmatrix}
      -\frac{215}{32} & 0 & 0 & 0 & 0 & 0 \\
      0 & -\frac{215}{32} & 0 & 0 & 0 & 0 \\
      3 & 0 & 1 & 0 & 0 & 0\\
      0 & 3 & 0 & 1 & 0 & 0\\
      \cos(2\theta_W)-\frac14 & 0 & 0 & 0 & 1& 0 \\
      0 & \cos(2\theta_W)-\frac14 & 0 & 0 & 0 & 1
     \end{pmatrix}\ .
\end{equation}
This matrix describes the mixing of QCD-like operators
and neutral weak electroweak operators and
determines the running of the Wilson coefficients.

The solution of the renormalization-group equation,
\begin{equation}\label{eq:RGE}
  \frac{d \vec{\will{}{}}}{d \ln \mu} 
         = \gamma^T(\mu) \cdot  \vec{\will{}{}} , 
\end{equation}
is straightforward,
and we find
\begin{eqnarray}
 \ten{\tilde I}{(2)}{q}(\mu) 
                      &=& \ten{\tilde I}{(2)}{q}(\Lambda)
                             \left(\frac{\alpha_s(\Lambda)}{\alpha_s(\mu)}\right)^{-\frac{215}{24\beta_0}} ,  \\
 \ten{I}{(2)}{q}(\mu)
                      &=& \left[ 3 \ten{\tilde I}{(2)}{q}(\Lambda)
                                   + \ten{I}{(2)}{q}(\Lambda) \right]
                                        \left(\frac{\alpha_s(\Lambda)}{\alpha_s(\mu)}\right)^{\frac{4}{3\beta_0}}
                                   -3\ten{\tilde I}{(2)}{q}(\Lambda)
                                        \left(\frac{\alpha_s(\Lambda)}{\alpha_s(\mu)}\right)^{-\frac{215}{24\beta_0}} ,  \allowdisplaybreaks \\
 \ten{I}{\prime q}{2}(\mu)
                      &=& \left[ \left(\cos(2\theta_W)-\frac14 \right)\ten{\tilde I}{(2)}{q}(\Lambda)
                                   + \ten{I}{\prime q}{2}(\Lambda)\right]
                                        \left(\frac{\alpha_s(\Lambda)}{\alpha_s(\mu)}\right)^{\frac{4}{3\beta_0}}\notag\\
                      &&      -  \left(\cos(2\theta_W)-\frac14\right)\ten{\tilde I}{(2)}{q}(\Lambda)
                                        \left(\frac{\alpha_s(\Lambda)}{\alpha_s(\mu)}\right)^{-\frac{215}{24\beta_0}} ,
\end{eqnarray}
with $I\in\{L,R\}$.
The scale dependencies of
$\coupL{\prime q}{1}$ and 
$\coupR{\prime q}{1}$ are given by
\begin{equation}
 \coupL{\prime q}{1}(\mu) = \coupL{\prime q}{1}(\Lambda) ,  \ \coupR{\prime q}{1}(\mu) = \coupR{\prime q}{1}(\Lambda)\ .
\end{equation}
With these relations it is not 
difficult to check the logarithmic terms
of the one-loop calculation including the logarithmic
corrections to every order in $\alpha_s \ln \Lambda/\mu$,
$\mu\sim m_t$.
\end{widetext}
\bibliography{Paper.bib}
\end{document}